# REVIEW

# Isotachophoresis applied to chemical reactions

C. Eid and J.G. Santiago

**ABSTRACT**

This review discusses research developments and applications of isotachophoresis (ITP) to the initiation, control, and acceleration of chemical reactions, emphasizing reactions involving biomolecular reactants such as nucleic acids, proteins, and live cells. ITP is a versatile technique which requires no specific geometric design or material, and is compatible with a wide range of microfluidic and automated platforms. Though ITP has traditionally been used as a purification and separation technique, recent years have seen its emergence as a method to automate and speed up chemical reactions. ITP has been used to demonstrate up to 14,000-fold acceleration of nucleic acid assays, and has been used to enhance lateral flow and other immunoassays, and even whole bacterial cell detection assays. We here classify these studies into two categories: homogeneous (all reactants in solution) and heterogeneous (at least one reactant immobilized on a solid surface) assay configurations. For each category, we review and describe physical modeling and scaling of ITP-aided reaction assays, and elucidate key principles in ITP assay design. We summarize experimental advances, and identify common threads and approaches which researchers have used to optimize assay performance. Lastly, we propose unaddressed challenges and opportunities that could further improve these applications of ITP.

## I. Introduction and Background

Although the term "isotachophoresis" was coined only 47 years ago, similar techniques have existed for nearly a century. In 1923, Kendall and Crittenden[1] described a technique to separate acids and metals, which they called the "ion migration method". After that, several studies in the 1930s and later described "moving boundary electrophoresis"[2] and "displacement electrophoresis",[3] processes nearly identical to isotachophoresis. It wasn't until 1970 that Haglund[4] introduced the term "isotachophoresis" (ITP), based on the fact that all ITP zones migrate at the same velocity at steady state ("isos" meaning equal and "takhos" meaning velocity in Greek). In the following years, ITP enjoyed a period of significant popularity, owing in part to the fact that it could be performed in capillaries larger than those used for capillary electrophoresis (CE).[5] The 1980s saw a decrease in ITP's popularity, but also the rise of a new area of application of ITP. Up to this point, ITP had mostly been limited to a preconcentration and separation technique. Then, in a number of publications led by researchers like Bocek[6,7] and Furst,[8,9] ITP applications were expanded to include the initiation and control of chemical reactions, particularly those involving enzymes and peptides. In those assays, ITP's high-resolving capability would be used to detect concentration changes in the enzyme-catalyzed mixture of reactants and products at several time points of the reaction.

ITP re-emerged in the spotlight in the 1990s through its coupling to CE. Researchers designed assays which incorporated ITP focusing followed by a disruption of ITP and initiation of CE separation. This combination provides dramatic increase in sensitivity while preserving the excellent separation efficiency of CE.[10-12] In recent years, ITP has found increased adoption in microfluidic formats, which leverage its various advantages (including self-sharpening zones, insensitivity to errors in injection or disturbances, and purification capabilities) in a wide array of applications. For readers interested in learning more about the history and development of ITP, we refer to several excellent electrophoresis and isotachophoresis reviews.[12-17] Indeed, today there are typically 40-50 papers published per year using ITP, the vast majority (>90%) of which being in microfluidic formats. These publications cover a wide range of applications, such as preconcentration of analytes prior to CE separation,[12,18,19] purification from complex samples,[20-25] analytical and computational modeling,[26-28] and fractionation of biological and chemical species.[29-31]

Unlike the majority of electrophoretic methods, ITP uses a two-buffer system consisting of a high-mobility leading electrolyte (LE) buffer and a low-mobility trailing electrolyte (TE) buffer. Sample ions with effective mobility magnitudes greater than the TE (in the TE zone) and less than the LE (in the LE zone) focus at an interface between these two co-ions.[32,33] Preconcentration factors of up to one-million have been achieved,[34] although 1,000 to 15,000-fold preconcentration is typically observed with more complex biological samples such as nucleic acids in blood.[22] The LE and TE zones have respectively high and low conductivity, and so a relatively high electric field is established in the TE zone and a low field in the LE zone. In accordance with the conservations of species and current, the TE and LE travel at the same rate. The strong electric field gradient establishes a self-sharpening and translating TE-to-LE interface which makes ITP robust to disturbances like pressure-driven flow, rough channel surfaces, and changes in channel geometry.

In Figure 1, we demonstrate qualitatively the self-sharpening feature of ITP. TE ions which diffuse into the LE zone experience significantly lower electric field, and are thus overtaken by neighboring

LE ions and fall back into their original TE zone. Conversely, higher mobility LE ions diffusing into the higher electric field TE zone are restored since they migrate faster than the TE. Importantly, TE and LE mobilities are chosen such that sample ions in the TE (LE) migrate faster (slower) than neighboring TE (LE) ions and are driven toward the TE-to-LE interface. See for example Khurana et al.[35] and Garcia et al.[36] for more detailed and quantitative descriptions (including models and experimental studies) of the diffusion- and dispersion-limited focusing dynamics of ITP sample ions.

ITP processes can conveniently be categorized as either peak-mode or plateau-mode. Peak-mode ITP is associated with sample ions present in trace concentrations. Such samples focus into the TE-to-LE interface region but there is insufficient time (and equivalently distance along the channel) for the sample ions to appreciably influence local ionic conductivity in the channel.[37] In peak-mode ITP, the sample species respond solely to the electric field established by the dynamics of the TE and LE. Importantly, multiple sample ions can co-focus within and significantly overlap within the same sharp ITP interface. In an approximate sense, well-focused sample ions accumulate into Gaussian peaks with continuously increasing area and significant spatial overlap. The focusing and relative positions of these peaks is determined solely by the electric field established by the TE and LE and the relative mobilities of the TE, the LE, and each sample species.

The second useful category for ITP is plateau-mode ITP. Above a certain threshold concentration (and duration of the ITP process), sample ions accumulate and reach a local maximum concentration. Here, multiple sample ions will reach respective maximum concentration and segregate into respective, multiple plateau-like zones of locally uniform (and constant) concentration. If there is a continuous influx of sample ions (e.g., from a reservoir), these plateaus increase in length in proportion to the amount of electrical charge run through the system.[38] For the rare case of ITP of fully-ionized species, this threshold is determined by the Kohlrausch regulating function (KRF).[39] For the common case of weak electrolytes (e.g., LE and TE solutions which are pH buffers), the threshold is governed by the Alberty[40] and Jovin[41] functions instead. Plateau-mode ITP has been leveraged for many applications, including separation and indirect detection of toxins, amino acids, and others.[42-44] Briefly, peak-mode ITP is well-suited for mixing and driving reaction kinetics due to co-focusing of trace sample ions into high-concentration, overlapping peaks; while plateau-mode ITP is better-suited for separation of species into distinct zones for the purpose of purification or identification.

In this review, we outline and discuss an emerging use and field of application of ITP: The initiation (via mixing), control, and acceleration of chemical reactions involving at least one ionic species. Accordingly, we will specifically consider applications of ITP wherein at least one reagent in a chemical reaction is focused at an ITP interface, and this focusing is used to control a chemical reaction involving that reagent and at least one other reagent. We first briefly summarize simple concepts of second-order reactions. We then review a series of papers in which ITP was used to preconcentrate and mix reagents, and describe mixing time scales for two adjoining zones. We then review papers using ITP to accelerate chemical reactions, and separately discuss homogeneous and heterogeneous assay configurations. For each class of application, we review and describe physical modeling and scaling of ITP-aided reaction assays and summarize relevant literature. In Table 1, we summarize the studies discussed in this review, classify these in a manner consistent with our discussions, and briefly mention their major contributions. In Table 2, we characterize the reactant species, kinetics, and performance of the reactions described in these studies. Lastly, we discuss unaddressed challenges and make recommendations for promising areas for future work.

## II. Earliest work involving ITP to control chemical reactions

A common theme in the first set of papers on ITP-aided reaction acceleration is the use of ITP to mix and control reactants in an ITP zone. We first provide a brief and simple scaling analysis to provide some physical context for this process. Unlike other types of microfluidic mixing using stirring or chaotic flows, mixing in ITP is typically accomplished via a deterministic electrophoretic process wherein one species is electromigrated into a region occupied by a second species. As mentioned above, migration velocity is the product of the electrophoretic mobility and the local electric field,

$$U_i = \mu_i E \tag{1}$$

Here, $U_i$ represents the velocity of a migrating species, $\mu_i$ is the local electrophoretic mobility (the sign of which indicates direction), and $E$ the local electric field. Consider the case of two analyte species, A and B, occupying two adjoining zones in a channel. For now, consider that both of these are present as trace species in a background of buffer ions. In such a case, their differential electrophoretic velocity can be quantified in terms of their effective mobilities. The two species mix when their different electrophoretic velocities cause relative motion toward each other. The time over which the two species would mix (i.e. overlap fully) scales as

$$t_{mix} \propto \frac{\delta_1}{U_2 - U_1} \propto \frac{\delta_1}{E(\mu_A - \mu_B)} \tag{2}$$

where $\delta_1$ denotes the width of smaller of the two zones, $E$ denotes the local electric field in zone 1, and $\mu_A$ and $\mu_B$ represent the electrophoretic mobilities of species A and B. As eq 2 shows, the mixing rate is influenced by the relative mobilities of the two species, the width of the zone, and the local electric field. The closer the two mobilities are to each other, the longer they will take to mix. We note that the latter scaling is also useful when one species is in ITP (plateau or peak mode) and the second has a mobility and initial position configured so that it will pass through the space occupied by the first. In such a case, the characteristic difference in velocity can be characterized as the difference between the ITP velocity (which is in turn the velocity of the LE co-ion) and the local electrophoretic velocity of the second species.

To our knowledge, the first demonstrated use of ITP to mix and initiate chemical reactions came in 2008 from scientists working at Wako Pure Chemical Industries, in a series of papers describing the development of the assay concept, its optimization, and its development into a commercial instrument. The result of this work, the µTASWako i30,[45] was the first commercially-available instrument that uses ITP.

In their first paper, Kawabata et al.[46] described an assay that they called the Electrokinetic Analyte Transport Assay. The assay leveraged ITP to focus a DNA-coupled antibody and increase its concentration while reacting with a target protein, α-fetoprotein (AFP). Conjugating the antibody with a DNA molecule increased its electrophoretic mobility and enabled the DNA-antibody to focus in ITP. The differential velocity between the ITP-focused DNA-antibody complex and the AFP (which was not focused in ITP) was used to initiate the reaction. The DNA-antibody complex reacted with AFP and Kawabata hypothesized that the ITP preconcentration of the DNA-antibody complex accelerated these reaction kinetics (neither quantitative data nor analysis supporting accelerated kinetics was provided) . Further, the reaction resulted in recruitment of unfocused AFP into ITP mode, increasing product concentration by up to 140-fold. Applied voltages where then reconfigured on their chip to initiate CE and separate the immune complex of interest from background fluorescent signal, as shown in Figure 2. The plastic microfluidic chip was designed as a straight channel with several branches to allow the introduction of the various buffers and reagents. Their LE and TE buffers contained Tris-HCl and Tris-HEPES, respectively, and additional components like polymers, albumin, salts, and surfactants to improve assay performance. They achieved a limit of detection of 5 pM with this assay, impressively nearly 2 orders of magnitude below clinically-relevant limits.

Simultaneously (the papers were published within the same week), Park et al.[47] published a paper on improving the reproducibility of the assay. They studied peak intensity and separation, and their dependence on "handoff time", the moment at which voltage switching causes the assay to transition from ITP stacking to CE separation. Interestingly, they found that changes in buffer concentration or small manufacturing defects in the devices caused noticeable variation in arrival times, which in turn affected handoff and adversely affected data quality. To combat this, Park introduced automated handoff and timing mechanisms which relied on computer monitoring of voltage, in order to adjust for external factors, and to achieve highly precise control of signal intensity and peak separation.

The final paper in this series was published in 2009, by Kagebayashi et al.[48] In it, they described the automated AFP-L3 assay, and the μTASWako i30 immunoanalyzer which evolved from the previous two papers. Kagebayashi et al. described its mechanism and characterized its performance. In addition to quantifying total AFP levels, they also incorporated an affinity-based separation step to simultaneously quantify the L3 isoform of AFP, AFP-L3. AFP-L3% is a clinically-relevant biomarker that is specific to malignant tumors and other pathologies.[49,50] By specifically binding to the L3 isoform through ITP preconcentration, the DNA-AFP-L3 immunocomplex separated from AFP-L1 isoform, allowing the quantitation of both isoforms using laser-induced fluorescence. They validated their assay in spiked serum samples, and achieved a limit of detection of 1 pM, with 2% coefficient of variation. Their test demonstrated good correlation with a commercially-available reference assay. The μTASWako i30 immunoanalyzer received FDA 510(k) clearance in 2011,[51] and remains commercially available as of this publication.

**III. ITP to preconcentrate, mix, and accelerate homogeneous reactions**

**III.a. Homogeneous reactions: theory and models**
In this section, we will briefly review analytical modeling and scaling of homogeneous chemical reactions (i.e. all reactants suspended in solution) using ITP. Standard second-order chemical reactions can be expressed as

$$A + B \underset{k_{off}}{\overset{k_{on}}{\rightleftarrows}} AB \qquad (3)$$

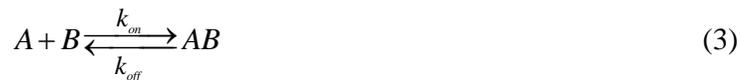

where $k_{on}$ and $k_{off}$ are the reaction on- and off-rate constants, respectively. The characteristic hybridization time scale at which half the limiting species (here, reactant B) can be expressed as

$$\tau_{std} = \frac{\ln 2}{k_{on} c_A^0} \quad (4)$$

where $c_A^0$ represents the initial concentration of reactant A.

Bercovici et al.[52] developed the first analytical model examining ITP-aided chemical reactions wherein both reacting species are focused in peak-mode ITP. This initial model assumed a perfectly overlapped Gaussian reactant peaks. They developed a volume-averaged set of first-order differential equations to describe conservation of species:

$$\begin{cases} \dfrac{dc_A}{dt} = \dfrac{Q_A^{TE}}{A} - \dfrac{1}{\delta}\dfrac{d\delta}{dt} - \dfrac{3}{\sqrt{\pi}} k_{on} c_A c_B + k_{off} c_{AB} \\ \dfrac{dc_B}{dt} = \dfrac{Q_B^{LE}}{A} - \dfrac{1}{\delta}\dfrac{d\delta}{dt} - \dfrac{3}{\sqrt{\pi}} k_{on} c_A c_B + k_{off} c_{AB} \\ \dfrac{dc_{AB}}{dt} = -\dfrac{1}{\delta}\dfrac{d\delta}{dt} + \dfrac{3}{\sqrt{\pi}} k_{on} c_A c_B - k_{off} c_{AB} \end{cases} \quad (5)$$

Here, $Q_A^{TE}$ and $Q_B^{LE}$ represent the influx of A and B from the zones wherein they were initially loaded (the TE for species A, and LE for species B). These rates are given by

$$Q_A^{TE} = \left(U_S^{TE} - U_{ITP}\right) A c_S^{ATE} = \left(\frac{\mu_A}{\mu_{TE}} - 1\right) p_{S,TE} U_{ITP} A \beta c_S^0$$

and

$$Q_B^{LE} = \left(U_{ITP} - U_S^{LE}\right) A c_S^{LE} = \left(1 - \frac{\mu_B}{\mu_{LE}}\right) U_{ITP} A c_S^0 = p_{S,LE} U_{ITP} A c_S^0$$

(6)

$U_{ITP}$ is the velocity of the ITP zone, $c_A^0$ and $c_B^0$ the respective reservoir or initial concentrations of species A and B, and $\beta$ is the ratio of TE ion concentrations in the adjusted TE and TE zones. For more on adjusted TE zones, we refer interested readers to Khurana et al.[35] and to Eid and Santiago[53]. $\delta$ is the width of the ITP zone, and is determined by the balance of dispersion effects (e.g., diffusion which acts to mix species and broaden the peak) and electromigration (which acts to sharpen the interface). In ideally, diffusion-limited conditions this width can be estimated as [54]

$$\delta_{theory} = \frac{RT}{F U_{ITP}} \left(\frac{\mu_{LE} \mu_{TE}}{\mu_{LE} - \mu_{TE}}\right) \quad (7)$$

where $R$ and is the universal gas constant, $T$ is the temperature, and $F$ is Faraday's constant. We note that in practice, the width of the ITP zone is not constant, but grows slightly over time.[35,52] Eq 6 also contains the so-called separabilities $p_{A,TE}$ and $p_{B,LE}$, which were first introduced by Bocek[55] and then further developed by Marshall.[56] For a generic sample ion $s$, separabilities are given by

$$p_{S,TE} = \frac{\mu_S}{\mu_{TE}} - 1$$
$$p_{S,LE} = 1 - \frac{\mu_S}{\mu_{LE}} \tag{8}$$

Separabilities quantify the relative mobilities of a species and the buffer it is in, and are useful in estimating the focusing rate of species.[53]

By assuming that one of the two reactants is in relative excess at the ITP interface, as well as an equilibrium constant which is low compared to the local concentration of the species in excess, an analytical solution to the system in eq 5 can be obtained. Under those circumstances, the above system of equations simplifies to a single ordinary differential equation, with the following exact and approximate solutions:

$$\bar{c}_{AB} = \frac{Q_B^{LE}}{\delta A}\left(t - 0.5\sqrt{\frac{\pi}{a}}e^{-at^2}\mathrm{erfi}\left(\sqrt{at}\right)\right) \cong \frac{Q_B^{LE}}{\delta A}t\left(1 - e^{-at^2}\right) \tag{9}$$

where $a = \frac{3}{2\sqrt{\pi}}\frac{Q_A^{TE}}{\delta A}$.

The simplest version of this volume-averaged unsteady product concentration can be compared to the value for a well-stirred reaction under similar assumptions, $c_{AB} \cong c_{Bo}\left(1 - e^{-c_{Ao}k_{on}t}\right)$. We see that the effect of ITP can be interpreted as a pseudo-first order reaction wherein the initial value of the excess reactant $c_A^0$ and the low-abundance concentration $c_B^0$ (which normally limits the maximum level of $c_{AB}$) are now functions directly proportional to time. Hence, this work revealed a new characteristic timescale for ITP-aided reaction kinetics, inversely proportional to square-root of initial concentration (compared to standard incubation, where the timescale is inversely proportional to initial concentration), as well as preconcentration due to ITP, and is given by

$$\tau_{ITP} = \sqrt{\frac{\ln 2}{\frac{Q_A^{TE}}{\delta A}k_{on}}} \tag{10}$$

Putting the two reaction timescales from eqs 4 and 10 together elucidates the effect of ITP on acceleration of reaction kinetics:

$$\frac{\tau_{std}}{\tau_{ITP}} \cong \sqrt{\ln 2} \sqrt{\frac{Q_A^{TE}}{\delta A k_{on} \left(c_A^0\right)^2}} \qquad (11)$$

They supported their model with experimental validation using a molecular beacon probe and oligo target. We summarize some of these results in Figure 3. ITP enhancement was more pronounced at lower reactant concentration (14,000-fold reduced reaction time at 500 pM target concentration), a regime in which reactions are governed by the off-rate. Though their work nominally focused on DNA hybridization, it is theoretically applicable to any ITP-aided reaction assay in which both reactants are preconcentrated in ITP.

Eid et al.[57] presented a modified version of Bercovici's model for cases in which only one species is focused in ITP. Namely, Eid considered loading of two reactants into the LE buffer, but where only one of them focused in ITP. This resulted in a reduction of ITP-enhanced reaction rate. Under these conditions, the net reaction rate of the not-yet-focused species within the LE zone may be comparable (on a moles per second basis) to that in the ITP zone, due to the significantly larger volume of the LE zone. They modeled the latter effect as a shrinking reactor, and introduced a dimensionless parameter $\lambda$ which incorporates several of the key variables which influence product formation

$$\lambda = \frac{L_0 k_{on} c_B^0}{U_{ITP}} \qquad (12)$$

Here $L_0$ is the length of the separation region in the channel.

All the models discussed above make the simplifying assumption that ITP zones have a Gaussian profile and were perfectly overlapped, which allows the use of volume-averaged concentrations. Garcia et al.[36] first showed that samples focused in ITP peak mode may exhibit species-specific and non-Gaussian/asymmetric profiles. In their work, they found that sample ion properties contributed greatly to dispersion and ITP peak shapes. In particular, ITP peaks wherein sample ions had mobilities near those of the TE or LE exhibited significant tailing into these respective zones and an associated asymmetry. Rubin et al.[58] presented a study focused on sample distribution within ITP zones, and the effect of these species specific distributions on reaction rates. They presented closed-form solutions for peak shapes and production rates for the case of ITP dynamics dominated by pure diffusion (e.g., no advective dispersion) and electromigration. To account for sample zone shape asymmetry, they defined an effective association rate constant of the form

$$k_{on}^{eff} = k_{on} k_{form} \qquad (13)$$

Here, $k_{form}$ depends on the relative sample, TE, and LE mobilities, and is given by

$$k_{form} = \frac{1}{\pi A \delta} \frac{1-(y_A + y_B)}{\cot(\pi y_A) + \cot(\pi y_B)} \quad (14)$$

$$y_A = \frac{1/\mu_{TE} - 1/\mu_A}{1/\mu_{TE} - 1/\mu_{LE}}, \quad y_B = \frac{1/\mu_{TE} - 1/\mu_B}{1/\mu_{TE} - 1/\mu_{LE}} \quad (15)$$

Interestingly, they found that reaction rate is not necessarily maximized when concentration profiles of two reacting species perfectly overlap, and instead varies depending on the relative mobilities of the species. As a result, production rates calculated while accounting for sample distribution are typically lower than those computed with volume-averaged concentration models.

Recently, Eid and Santiago[53] considered the design parameters that govern the performance of peak-mode ITP assays. They incorporated the results of Rubin's more comprehensive species overlap model into their analysis. This analysis showed that for reaction times longer than the characteristic ITP reaction time scale $\tau$, the number of molecules of product AB formed depends solely on the relative influx rates of reactants A and B, and defined a dimensionless production rate such that

$$\hat{N}_{AB}^j (t \gg \tau) \equiv \frac{N_{AB}^j}{Q_B^j t} = \frac{Q_A^j}{Q_A^j + Q_B^j} \quad (16)$$

Here, $j$ represents the initial loading buffer (LE or TE buffers), and $N_{AB}$ represents the number of AB molecules formed. For the case in which B is in relative abundance to A, the above equation simplifies to unity, implying that after a short transient associated with kinetic rates, production rate is limited by and equal to the net influx rate of the rate-limiting species (the species present in locally higher concentration). This finding is consistent with what Shintaku et al.[59] reported in their examination of ITP-aided acceleration of bead-based reactions. Furthermore, they considered the effect of initial sample placement on production rates. They defined $\varepsilon$, a ratio of product formation when both reactants are initially loaded into the TE versus when both are loaded in the LE buffer,

$$\varepsilon \equiv \frac{N_{AB}^{TE}}{N_{AB}^{LE}} \quad (17)$$

For long reaction times, they found that $\varepsilon$ depends solely on $\phi_A$, the influx ratio of the limiting species,

$$\varepsilon(t \gg \tau) \cong \phi_A = \frac{p_{A,TE}}{p_{A,LE}} \beta \quad (18)$$

Eid and Santiago concluded that for sufficient reaction times, the optimal production rate of species AB in an ITP-aided reaction assay is obtained by simply maximizing the influx rate of the rate-limiting species.

**III.b. Homogeneous reactions: experimental studies and assays**

*III.b.1. Homogeneous nucleic acid hybridization assays*
We here consider nucleic acid hybridization reactions. We term these as "homogenous" when they involve at single-stranded nucleic acid species which are in solution (i.e. not attached to a substrate). Such assays are attractive due to their simple design and implementation. However, excess reactant removal and clean-up steps can be more difficult to incorporate into the workflow (e.g., relative to heterogeneous reactions). Importantly, multiplexing is much more difficult to achieve in this format. The first discussion of using ITP to mix reagents in the context of ssDNA hybridization came from Goet et al.[60] in 2009. This analysis came in the greater context of using ITP to bring sample zones into well-controlled contact. However, Goet only discussed this possibility and did not experimentally demonstrate the concept.

To our knowledge, Persat and Santiago[61] were the first to experimentally demonstrate ITP-aided hybridization of nucleic acids. They used ITP and molecular beacons to selectively profile among seven microRNA (miRNA) species in total RNA samples from human liver and kidney. This is also the first published quantitative demonstration of high specificity reaction using ITP; and signal selectivity to target miR-26 (22 nt), compared with its relative, miR-126 (22 nt), and its precursor, pre-mir-26a (77 nt). Molecular beacons are single-stranded DNA (ssDNA) probes with a unique hairpin structure that places a fluorophore and quencher in close proximity.[62] Upon binding to a specific target, the structure of the probe is disrupted, separating the fluorophore from its quencher, and resulting in increased fluorescent signal. Persat and Santiago developed a multistage assay wherein the channel contained three discrete regions with varying amounts of sieving matrix (polyvinylpyrrolidone, PVP), magnesium chloride, and denaturant. In the first region, all RNA molecules in the total RNA sample were preconcentrated in ITP. The second region contained a high concentration of sieving matrix in order to defocus large RNA and selectively retain miRNA. The third region applied stringent conditions to promote specific hybridization between the molecular beacons and miRNA target. Figure 4 shows a schematic of their multi-stage assay. Persat verified both sequence- and size-selectivity of this assay; the former by titrating with mismatched miRNA, and the latter by titrating with larger precursor miRNA. They demonstrated initial biological relevance of this technique by achieving a 10 pM limit of detection of miR-122 in kidney and liver total RNA samples.

Bercovici et al.[63] combined ITP-aided reactions with molecular beacons, but used it to detect a significantly larger target, 16S rRNA from bacteria in cultures and patient urine samples. To our knowledge, this is the first quantitative demonstration of reaction acceleration using ITP in addition to the first quantitative theory and validation thereof. They successfully demonstrated the applicability of this approach in real patient samples at clinically-relevant levels. Another significant contribution of the latter work is the design of a photomultiplier tube (PMT) system for high sensitivity fluorescent signal quantification. The assay achieved a limit of detection of $10^6$ cfu/mL, or 30 pM, of *E. Coli* in human urine samples. This work and that of Persat and Santiago[61] demonstrated that molecular beacons can be used for selective assays; but also that ITP assays with molecular beacon based detection sacrifice sensitivity and dynamic range (due to large background signal of unreacted MBs). These limitations highlighted the need for removal (e.g., physical separation) of background signal following hybridization, and subsequent work in the area has devised and demonstrated various solutions to this issue.

Bahga et al.[64] introduced a homogeneous ITP reaction assay for removing excess background signal of unreacted molecular beacons. They designed an assay in which ITP-aided DNA hybridization was coupled to the high-resolving power of CE using bidirectional ITP (see review by Bahga and Santiago[12]). Following ITP-aided hybridization, CE was triggered, and unbound molecular beacons separated from the larger (lower mobility) beacon-target complex. They successfully demonstrated sequence-specific detection of a 39 nt ssDNA target, with a 3 pM limit of detection. Though Bahga et al. improved sensitivity of ITP and molecular beacon assays with this approach, the use of CE resulted in more dispersed signal peaks, limiting sensitivity and downstream analysis of reaction products. Another drawback of this approach is the complexity of assay chemistry needed for effective bidirectional ITP.

A few months later, Eid et al.[65] introduced a new method for homogeneous post-hybridization clean-up. Their multistage approach used ITP to accelerate hybridization between a 26 nt linear ssDNA probe and 149 nt ssDNA target, and an ionic spacer to subsequently separate reaction products. In a similar vein to Persat et al.,[61] the first stage focused all probe and target, promoting rapid mixing and hybridization. The second stage contained high-concentration sieving matrix, which allowed the ionic spacer to overtake and separate the unbound probes from the slower probe-target complex. Figure 5 shows the different stages of the reaction-separation assay. Eid et al.[65] used 20 mM HEPES as TE, 1 mM MOPS as spacer, 1.8% hydroxyethyl cellulose (HEC) as sieving matrix. This resulted in two focused ITP peaks, one at the LE-spacer interface, and one at the spacer-TE interface. They demonstrated the advantage of this approach by achieving a 220 fM limit of detection in 10 min, with a 3.5 decade dynamic range. This technique is fairly simple and flexible, and produces two ITP-focused peaks; this improves signal and is compatible with downstream manipulation. However, the method lacks high selectivity, and requires sieving matrix and spacer optimization for targets of different sizes.

*III.b.2. Bead-based homogeneous DNA assays*

Beads can be focused into ITP zones and used to achieve pseudo-homogenous reactions between species in solution and randomly dispersed beads. Beads offer the advantage of inherent multiplexing (e.g., by coding beads). Shintaku et al.[59] leveraged ITP to co-focus target DNA with DNA-conjugated beads in order to strongly accelerate multiplexed DNA hybridization reactions. They conjugated 6.5 μm polystyrene beads with ssDNA probe sequences corresponding to ten different target oligos. Since fluorescent quantitation was performed using a Luminex 200 instrument, the need for additional signal removal methods was obviated. They also developed a model to describe the reaction kinetics of bead-based hybridization with and without ITP. Interestingly, Shintaku described a quasi-equilibrium at high target concentrations between target influx and consumption within an ITP zone, and Eid and Santiago[53] would later expand on this finding (see Section III.a). Their 20 min assay achieved comparable sensitivities to a standard 20 h standard hybridization assay, and 5-fold higher sensitivity when compared with 30 min of standard hybridization. Furthermore, their multiplexed assay (ten target species) showed similar specificity as the standard bead assay, as measured by a so-called specificity index, which was defined as the ratio of specific signal to the highest nonspecific signal.

*III.b.3. Homogeneous immunoassays aided by ITP*

Unlike nucleic acids, which have relatively high and largely size-independent mobilities in free solution,[66] protein mobilities cover a wide range of mobilities and solubilities, and are difficult to model and predict *a priori*. Additionally, proteins are highly sensitive to pH levels and other environmental factors. As a result, researchers seeking to design ITP reaction assays with protein targets have either had to use well-characterized and high-mobility proteins, or devise clever workarounds to focus and detect proteins (e.g., such as the aforementioned conjugation of proteins with a DNA molecule by Wako[46]).

Eid et al.[57] devised a workaround, using high-mobility modified aptamers to bind to and recruit typically non-focusing C-reactive protein (CRP) into ITP. The modified aptamers are called SOMAmers (Slow Off-Rate Modified Aptamer), and unlike traditional aptamers that are DNA or RNA oligonucleotides, SOMAmers have modified bases to increase their hydrophobicity.[67,68] Eid here loaded both the protein and fluorescently-labeled SOMAmers into LE buffer. The ITP chemistry was chosen so as to preconcentrate the relatively high mobility SOMAmers and the SOMAmer/protein complex but not the unbound protein target. Hence ITP preconcentrated SOMAmers (increasing reaction rate) and simultaneously recruited proteins into the ITP zone upon binding. They extended the approach introduced by Eid and colleagues[65] combining ITP and ionic spacers to separate unbound SOMAmers from SOMAmer-protein complex to remove excess background signal. In clean buffer, this technique achieved a limit of detection of 2 nM, well within the clinically-relevant range of CRP. However, when extended to 20-fold diluted serum, assay sensitivity suffered, decreasing by an order of magnitude. Eid hypothesized this was due to the complexity of serum, due to its high protein content and associated nonspecific binding. The presence of many ionic species which act as native spacer molecules may have also contributed.

*III.b.4. Whole cell ITP reaction assays*

In recent years, ITP has been used to accelerate reactions involving whole dispersed cells. Schwartz and Bercovici[69] used positively-charged antimicrobial peptides (AMPs) to detect (but not identify) the presence of whole bacterial cells from a water sample spiked with *E. coli*. The cell-AMPs reaction is not specific to *E. coli* strain but demonstrated a new whole cell/reactant application. While previous studies used ITP to focus and detect pre-labeled (i.e., pre-reacted) bacterial cells in river water samples,[70,71] this was the first to perform in-line labeling reaction, detection, and separation. Here, Schwartz used cationic ITP to focus the AMPs and used pressure-driven counterflow to hold them the high-concentration ITP zone stationary for over 1 h. Figure 6 shows the schematic of this assay and an image of the stationary ITP peak. Meanwhile, ITP and pressure-driven flow acted unidirectionally on *E. coli* bacterial cells, resulting in continuous flow of bacterial cells reacting with and labeled by the stationary (in the lab frame) AMPs. AMP concentration was limited by the tendency of positively charged AMPs to bind to negatively-charged glass channel walls. The assay was capable of detecting $10^4$ cfu/mL over the 1 h detection window, which is on the upper limit of relevant concentrations, while using 100-fold fewer reagents compared to conventional AMP techniques. Schwartz and Bercovici hypothesized that an additional one or two orders of magnitude improvement in sensitivity may be possible using parallel channels for increased throughput.

Phung et al.[72] leveraged the use of counterflow pressure as well as an ionic spacer and sieving matrix to perform an ITP-aided fluorescence in situ hybridization (FISH) assay on bacterial cells. In the first step of their assay, bacterial cells and fluorescently labeled oligonucleotide probes are co-focused in peak-mode ITP. The nature of their synthetic cDNA hybridization probes makes this, to our knowledge, the first example of a high specificity reaction between cells and a reactant. Phung used counter flow to immobilize the ITP zone and prolong the sequence-specific hybridization process. In a second stage of the assay, a sieving matrix was used to separate stained cells from excess probe. Probes specific to target species *E. coli* and *P. aeruginosa* were respectively labeled with FAM (excitation wavelength of 488 nm) and Cy5 (excitation wavelength of 635 nm). Similar to Eid et al.,[65] Phung used 1.8% HEC as sieving matrix, but used MES as an ionic spacer to separate stained cells from excess probe. They tested the selectivity of their technique by testing probes designed for different bacterial species, and found that selectivity heavily depends on probe design and the number of matched nucleotides with the bacterial target. The in-line ITP-FISH assay required 30 min to complete, a significant improvement over the 2.5 h typical of conventional FISH assays. However, the performance suffered due to reduced staining efficiency (approximately 50% of the 2.5 h conventional FISH offline hybridization reaction), and the assay achieved a limit of detection of 6 x $10^4$ cells/mL. While this concentration range excludes several interesting applications, ITP-FISH is a faster, more convenient, and automation-friendly alternative to conventional FISH assays.

*III.b.5. Coupling ITP with enzymatic reactions*

Although not yet demonstrated, there have been two interesting accomplishments toward integration of ITP with enzymatic processes. The first was a paper published by Borysiak et al.[20] in which loop-mediated isothermal amplification (LAMP) was used in conjunction with ITP. ITP was first used to extract and purify DNA from E. coli after proteinase K-based lysis of diluted whole milk samples. ITP was then deactivated and increased temperature was used to manipulate air pressure and fluid flow, and divert the ITP-focused DNA into a reaction chamber with dried LAMP reagents for on-chip amplification. Borysiak achieved 100-fold lower limit of detection compared with standard tube-based LAMP without ITP purification. More recently, Eid and Santiago[23] demonstrated the compatibility of ITP with another isothermal amplification method, recombinase polymerase amplification (RPA), as well as with alkaline and proteinase-K lysis. ITP was used to extract and preconcentrate genomic DNA from inactivated Gram-positive *L. monocytogenes* cells spiked into whole blood. The extracted DNA was then transferred directly to a tube containing RPA master mix. In the supplementary information of this reference, Eid further demonstrated preliminary results of a more integrated assay in which an electric field was used to migrate an ITP DNA sample zone into an LE reservoir containing RPA master mix and primers. ITP was then deactivated and on-chip isothermal amplification and detection was initiated. In both assays, ITP enabled isothermal amplification with minimal (10-fold or less) dilution of the original complex sample.

## IV. ITP to preconcentrate, mix, and accelerate heterogeneous reactions

### IV.a. Heterogenous reactions: theory and models

In heterogeneous assays, one or more of the reactants is immobilized on a solid (stationary) surface. Karsenty et al.[73] developed the first analytical model for surface-based hybridization assays using ITP. They considered the case of a single target species migrating over a finite segment of functionalized magnetic beads immobilized using an external magnet. Through scaling analysis, they showed that the problem can be assumed to be one-dimensional, meaning that the analyte concentration at each point $x$ along the channel can be effectively represented by its area-averaged concentration. Karsenty also showed that the spatial distribution of target ions remains unchanged as they react with the immobilized probes. They developed the following one-dimensional ordinary differential equation to describe the concentration of surface probes bound by target molecules

$$\frac{db(t)}{dt} = \frac{b_m}{\tau_{on}} - \frac{b(t)}{\tau_R} \tag{19}$$

where

$$\tau_R^{-1} = \tau_{on}^{-1} + \tau_{off}^{-1}; \quad \tau_{on}^{-1} = \alpha c_0 k_{on}; \quad \tau_{off}^{-1} = k_{off} \tag{20}$$

Here, $\tau_R, \tau_{on}$, and $\tau_{off}$ are the timescales that govern reaction, association, and dissociation rates, respectively. $b_m$ is the initial (total) number of available probe molecules, and $\alpha$ is ITP preconcentration factor. The duration over which the ITP-focused target overlaps with the stationary probe molecules is given by $\tau_{ITP} = \delta/U_{ITP}$, where $\delta$ is the width of the ITP peak. Karsenty defines an enhancement ratio of ITP-aided surface reactions compared with standard flow reactions,

$$R = \alpha \frac{k_{off} + c_0 k_{on}}{k_{off} + \alpha c_0 k_{on}} \frac{1 - \exp\left(-\left(k_{off} + \alpha c_0 k_{on}\right)\tau_{ITP}\right)}{1 - \exp\left(-\left(k_{off} + c_0 k_{on}\right)\tau_{tot}\right)} \tag{21}$$

which, at low concentrations, simplifies to

$$R \approx \alpha \frac{\tau_{ITP}}{\tau_{tot}} \qquad (22)$$

The above results indicate that at low concentrations, ITP-based enhancement depends on the preconcentration factor as well as the fraction of the total assay time in which the two reacting species overlap through ITP. At higher concentrations, the reaction rate, even without ITP, is sufficiently high to saturate the probes, and the enhancement ratio decays until it reaches unity.

The above model has been adapted by several subsequent works since the original publication. Han et al.,[74] in their work on ITP-aided acceleration of microarray assays, provided a similar analysis of ITP-aided surface hybridization. They define the fraction of surface probes hybridized following ITP reaction by

$$h_{ITP} = \frac{\alpha c_0 k_{on}}{\alpha c_o k_{on} + k_{off}} \left(1 - \exp\left(-\left(\alpha c_0 k_{on} + k_{off}\right)\tau_{ITP}\right)\right) \qquad (23)$$

Han et al. also identified the Damkohler number as an important dimensionless parameter governing these reactions. Damkohler number, $Da$, is the ratio of reaction rate and diffusion rate, and can be used to determine whether a system is reaction- or diffusion-limited. Typical ITP assays have $Da << 1$, indicating that reaction rates are much lower than diffusion, and thus these assays are reaction-limited. Moghadam et al. also used the analysis of Karsenty[73] in their work on ITP-aided lateral flow assays. They too found that their assay was reaction-limited, and their limit of detection was governed by the preconcentration and off-rate constant, similarly to Bercovici et al.[52]

Recently, Paratore et al.[75] expanded on the works of Karsenty and Han, and further investigated surface-based ITP reaction kinetics. They defined three operation modes for ITP: pass-over ITP (PO-ITP), stop and diffuse ITP (SD-ITP), and counterflow ITP (CF-ITP). The first of the three is the mode used in Karsenty's and Han's previous works, and consists of focused target passing over a reaction site containing immobilized probes, and reacting during a brief period of spatial overlap. In SD-ITP, once the focused target reaches the reaction site, the electric field was turned off, and the target diffused while reacting over the reaction site. To characterize reaction kinetics in this mode, the authors defined two characteristic time scales for depletion and axial diffusion, respectively:

$$\tau_{dep} = \frac{H}{b_m k_{on}} \qquad \tau_{D_x} = \frac{\delta^2}{D} \qquad (24)$$

Here, $H$ is the channel height. By assuming that the sample is well-mixed along the channel height, that diffusion along the y-axis is faster than the reaction rate, that the width of the reaction site is wider than the ITP zone width, and that the number of target molecules is significantly lower than available reaction sites, an analytical expression for fraction of bound surface molecules, equivalent to eq 23, was obtained:

$$h_{SD-ITP}(t) = k_{on}\tau_{dep}\left[\chi - \left(32\frac{t}{\tau_{D_x}}+1\right)^{0.5}\exp\left(-\frac{t}{\tau_{dep}}\right)\right]\alpha c_0 \qquad (25)$$

where

$$\chi = \left[1-erf\left(\sqrt{\frac{\tau_{D_x}}{32\tau_{dep}}}\right)\right]\sqrt{\frac{\pi\tau_{D_x}}{32\tau_{dep}}}\exp\left(\sqrt{\frac{\tau_{D_x}}{32\tau_{dep}}}\right) \qquad (26)$$

For sufficiently long times, longer than either characteristic time scales, the solution to eq 25 approaches a steady state of $\chi\alpha c_0$. The third and final mode of ITP they discussed is counterflow ITP (CF-ITP), in which the ITP peak was held steady over the reaction site by applying counterflow pressure. There, the fraction of surface molecules bound is given by

$$h_{CF-ITP}(t) = k_{on}\left(\sqrt{\frac{\pi}{8}}\alpha\frac{\tau_{dep}}{\tau_{ITP}}c_0\right)t \qquad (27)$$

Paratore's model revealed that SD-ITP and CF-ITP result in significantly more reaction acceleration compared with PO-ITP. SD-ITP is highly dependent on the diffusion behavior of the focused species, which can be improved by increasing viscosity. On the other hand, CF-ITP is dependent on the ratio of ITP accumulation and species depletion time scales, which can be controlled by optimizing ITP focusing conditions.

Shkolnikov and Santiago[76] developed a model for ITP coupled with affinity chromatography (AC), wherein target molecules are focused in ITP and probes are immobilized in an affinity capture region. Their analytical model they developed was broad and comprehensive. It contained fully-reversible second-order reaction kinetics, allowed for spatially variable bound substrate distribution, and accounted for zone dynamics of ITP-focused analytes. They found that key parameters like probe and target concentrations, forward and reverse reaction constants, and others collapsed into three dimensionless parameters that governed the different limiting regimes in the ITP-AC problem: the Damkohler number ($Da$), the ratio of peak target concentration scaled by the initial probe concentration ($\hat{c}_T$), and a dimensionless equilibrium constant ($\hat{K}_D$),

$$\hat{c}_T = \frac{a\sqrt{2\pi}}{c_{P,0}} \quad Da = \frac{\sigma k_{off} c_{P,0}}{U_{ITP}} \quad \hat{K}_D = \frac{k_{off}}{k_{on} c_{P,0}} \tag{28}$$

Here, $a$ is the maximum concentration of the Gaussian distribution, $\sigma$ its standard deviation, and $c_{P,0}$ the initial probe concentration. Shkolnikov found that increasing concentration via ITP can proportionally reduce capture times and required capture lengths across different $Da$ regimes, thus increasing column utilization compared with traditional AC assays. They also studied capture dynamics in a variety of interesting limiting regimes. For example, they found that scaled capture time reaches an asymptotic value for the low $Da$ regime, but increases linearly with $Da$ for high $Da$ values. The former limit is where capture time is limited by the advective time associated with the electromigration of the species in entering the capture region; while the latter limit is governed by the time-to-completion of the reaction (as determined by the species with the higher local concentration). They also found that the spatial resolution of ITP-AC purification scales proportionally with time for the case where the AC capture molecule is present at higher concentration.

**IV.b. Heterogeneous reactions: Experimental studies and assays**

*IV.b.1. Functionalized gel or porous regions and affinity chromatography type ITP assays*
Garcia-Schwarz and Santiago[77,78] developed a two-stage assay that used ITP to enhance hybridization and a photopatterned functionalized gel to remove excess reactant. The first of their two publications on this subject[77] is notable for its integration of serial ITP reactions with gel capture. ITP was first used to enhance the reaction kinetics between miRNA targets and fluorescently-labeled complementary reporters. In the second stage of the assay, the ITP zone migrated towards a photopatterned gel region which was functionalized with probes complementary to the reporters. Unused reporters bound to the probes and stayed behind as the reporter-target complex migrates through the gel. The resulting purified ITP zone was then fluorescently quantified to determine target concentration. Garcia-Schwarz demonstrated the selectivity of this technique for mature over precursor miRNA that are larger but contain the mature miRNA sequence. They achieved a 1 pM limit of detection with a 4 order of magnitude dynamic range using a linear DNA probe. In the second paper,[78] Garcia-Schwarz and Santiago built on their previous work by exploring specifically increases in assay specificity. They demonstrated single-nucleotide specificity with an assay which preferentially detected let-7a over other members of the let-7 microRNA family. Furthermore, they explored improved probe designs to enhance thermodynamic and kinetic specificity, and used locked nucleic acid (LNA) probes as well as a hairpin-structured reporter. The tradeoff for this enhanced specificity was reduced sensitivity and dynamic range, as they sacrificed an order of magnitude in the former and two orders of magnitude in the latter. Finally, they successfully validated this approach using total RNA samples, and demonstrated comparable results to PCR quantification. One limitation of this method of signal removal is the laborious experimental preparation required to pattern and functionalize the gel, and another is the challenge of performing multiple experiments on a single device. However, the demonstration of single-nucleotide specificity, a key requirement in many nucleic acid applications, was an important achievement showing the significant potential of ITP assays.

In the second of their two-part work, Shkolnikov and Santiago[76,79] presented a technique that coupled ITP preconcentration with affinity chromatography for sequence-specific capture and purification of target DNA molecules. The found that the three key parameters which they identified in their first paper ($Da$, $\hat{c}_T$, and $\hat{K}_D$) (and section IV.a above) showed remarkable consistency when compared with experimental results, accurately predicting the spatiotemporal behavior of the DNA. In addition to their analytical contributions (described in the previous section), they synthesized and functionalized a custom porous polymer monolith (PPM), made of poly(glycidyl methacrylate-*co*-ethylene dimetharcylate). They then functionalized this monolith with probe oligos, and characterized its performance. The PPM had little non-specific binding, was nonsieving, and compatible with ITP. Schkolnikov demonstrated the capability of this approach, purifying 25 nt Cy5-labeled target DNA from 10,000-fold more abundant background DNA in less than a minute, and their technique had a 4 order of magnitude dynamic range.

*IV.b.2. Surface-based DNA hybridization assays*
Karsenty et al.[73] were the first to experimentally demonstrate ITP-aided speed-up of a simple surface-based hybridization reaction. They designed a 3 min assay in which they used an external magnet to attract and immobilize molecular beacon-conjugated paramagnetic beads on a designated location on the microfluidic channel. To facilitate the capture of beads, the microfluidic channel contained a 100 x 30 µm trench. They then initiated ITP to transport sample containing target molecules over the prefunctionalized surface, allowing rapid hybridization over only 4 s of incubation. Contaminants and unbound target molecules continued migrating downstream, resulting in a single-step reaction and purification assay. They achieved a limit of detection of 1 nM, a 100-fold improvement in sensitivity over comparable continuous-flow surface hybridization assays. Interestingly, the improvement in sensitivity underperformed theoretical predictions which they attributed to in part to shorter reaction times for ITP. This initial feasibility type study used a single probe and a single target species, and did not explore specificity and cross-reactivity with interferents.

Han et al.[74] were the first to demonstrate multiplexed ITP-accelerated heterogenous reactions. They applied ITP to focus nucleic acids and accelerate a DNA hybridization microarray assay. A key feature of DNA array technologies is their massive multiplexing capacity, capable of detecting thousands of targets simultaneously.[80,81] Up until the work of Han, ITP studies were limited to the detection of one or two targets. Han detected 20 target sequences using 60 microarray spots. To achieve this, Han et al. designed a PDMS microfluidic superstructure containing a single channel that they bonded to a commercially available microarray on a glass substrate, exposing 60 spots to the path of the ITP zone. The resulting device was 40 µm deep, 8 cm long, and 500 µm wide, except for a short constriction that had width of 200 µm. The electric field was varied to optimize both ITP preconcentration and DNA hybridization (Figure 7). In the first part of the assay, they applied high electric field to accumulate all the ssDNA targets. Once the ITP zone reached the constriction, the electric field was turned off to allow target to redistribute through diffusion and minimize Joule heating effects. Finally, in the hybridization step, they applied low electric field values to avoid electrokinetic instabilities[37,82] as the sample migrated over the immobilized array spots. The 30 min assay achieved a 100 fM limit of detection with a dynamic range of 4 orders of magnitude. Compared with a conventional 15 h microarray hybridization assay, the 30 min ITP assay also demonstrated 8-fold increase in sensitivity. Furthermore, this improvement in sensitivity was achieved without any loss in specificity, as measured using the specificity index introduced earlier by Shintaku et al.[59] This work demonstrated the potential for ITP to integrate into DNA microarrays and other multiplexed surface reactions.

*IV.b.3. Heterogeneous immunoassays aided by ITP*

To the best of our knowledge, there have been two research groups (three studies) which used ITP to enhance heterogeneous immunoassays. In the first, Khnouf et al.[83] extended ITP-aided surface reaction assays to immunoassays using two different approaches for antibody immobilization. The first approach consisted of immobilizing antibody-coated magnetic beads at a specified region in the microchannel. In the second, they directly immobilized antibodies on a functionalized region of the channel. They designed a PMMA device that enabled electrokinetic injection in order to control the amount of sample injected. Following electrokinetic injection, the ITP-focused sample, containing the protein target, here bovine serum albumin (BSA), enters the reaction regions and binds to the immobilized antibodies. They empirically estimated a 100-fold preconcentration factor for the target protein. For both assay formats, using ITP to preconcentrate sample lead to an increase in sensitivity. Khnouf circumvented the issue of low protein mobility by selecting BSA, a high mobility protein, as their target, and labeling it with the even higher-mobility Alexa Fluor 488.[84] Despite this limitation, this study showed the feasibility of incorporating ITP in heterogeneous immunoassays.

Moghadam et al.[85] further demonstrated the applicability of ITP to one of the most widely-used immunoassay formats, lateral flow assays (LFA). Despite their convenience and popularity for qualitative applications, LFAs are often constrained by poor detection limits. Using ITP preconcentration, Moghadam demonstrated 2 orders of magnitude improvement in sensitivity in detecting IgG from a clean buffer. In conventional LFAs, samples migrate solely under capillary action, whereas they migrate electrophoretically in ITP-LFA (Figure 8). In addition to reaction enhancement achieved by ITP, Moghadam estimated an extraction efficiency of nearly 30%, more than an order of magnitude improvement over conventional LFA. Additionally, Moghadam offered significant insight into the general design of ITP-LFA. Capture efficiency in conventional LFA depends on the normalized initial target concentration. In ITP-LFA, however, capture efficiency also depends on preconcentration factor, time, and dissociation rate constant. As a result, optimal assay performance can be achieved by carefully designing buffer chemistry and device geometry.

Most recently, Paratore et al.[75] developed an assay in which they used different modes of ITP to accelerate surface-based immunoassays. They guided paramagnetic beads coated with anti-GFP antibodies to a desired location in the microfluidic channel, similar to Karsenty et al. They used enhanced GFP (EGFP) as a model protein, and demonstrated three separate operation modes of ITP: pass-over ITP (PO-ITP), stop and diffuse ITP (SD-ITP), and counterflow ITP (CF-ITP). For all assays, the total assay time was nearly 30 min, only 6 of which were for antibody-antigen binding. They found that SD-ITP, in which the electric field was turned off when the focused protein reached the reaction site, and CF-ITP, in which counterflow was applied to keep the focused proteins over the reaction site, resulted in significant decrease in LoD as compared with standard mode SD-ITP. Paratore reduced the LoD from 300 pM (for a standard flow immunoassay) to 220 fM using CF-ITP. They noted the influence of pH and salt concentration on the performance of the ITP immunoassay, and highlighted the heterogeneity of target proteins, which in turn requires careful optimization for each target. Despite the typical issues associated with ITP immunoassays (e.g. impact of extreme pH values and salt concentrations on focusing dynamics and reaction kinetics), Paratore's results are promising for the integration of ITP with protein detection assays.

## V. Conclusions and recommendations

We reviewed a number of theoretical and experimental studies wherein ITP focusing and mixing is used to initiate and accelerate chemical reactions. ITP preconcentration has been shown to increase reaction rate by up to 14,000-fold. By selectively focusing species within a designed electrophoretic mobility range, ITP can be used to remove contaminants and interfering species prior to a reaction, increase concentration during the reaction, and remove background signal or excess species following a reaction. Assays using ITP have demonstrated the ability to seamlessly integrate with various detection methods and other downstream assay steps. These assays leverage a variety of analytical chemistry tools including microfabrication, electric field control, a variety of cDNA-type probes, aptamer-type probes, antibodies, sieving matrices, spacer ions, denaturing conditions, microarrays and associated scanners, enzymatic reactions, and multispectral emissions and color detection. Downstream assays shown to be compatible with ITP include PCR, Luminex, and Bioanalyzer quantitation. Together, these studies demonstrate ITP's versatility and potential to control, enhance and accelerate chemical reactions. Furthermore, the continued near-decade existence of a commercial ITP-based immunoanalyzer suggests ITP's potential beyond the academic realm.

There remains several areas of unaddressed challenges and opportunities. There have been only a handful of immunoassays using ITP, and those have been limited to a few, particularly well-characterized proteins. Though the heterogeneity of protein mobilities, solubilities, and sensitivity to pH and ionic strength presents a significant challenge, the opportunities to incorporate ITP into proteomic analysis are significant and worth exploring. Integrating ITP with enzymatic processes, particularly amplification techniques like PCR and RPA, is also a significant opportunity for future ITP assays. Another area of opportunity is increasing the level of multiplexing in ITP-aided reactions from 20 species to thousands, and increasing throughput by analyzing and controlling samples in parallel.

The studies up to this point have been fairly limited in the input amount of sample that is processed. Increasing channel dimensions would potentially increase sensitivity and allow the processing and detection of rare species like viral DNA and circulating tumor cells. It may also encourage more widespread adoption of the technique. In particular, further incorporating ITP reaction process with complex sample extraction and purification (e.g. from blood, urine, tumor tissue, etc.) and with next-generation sequencing library preparation and target enrichment protocols would address key needs of those techniques (long incubation times, need for several washes). Such integration has immense potential.

**Table 1.** Summary of the twenty-three studies that applied ITP to chemical reactions.

| Authors | Year | Primary Contribution |
|---|---|---|
| **ITP to mix and control chemical reactions** | | |
| Kawabata et al.[46] | 2008 | First work to use ITP to speed-up chemical reactions |
| Park et al.[47] | 2008 | |
| Kagebayashi et al.[48] | 2009 | First commercial system to use ITP |
| **ITP to preconcentrate, mix, and accelerate homogeneous reactions** | | |
| Persat and Santiago[61] | 2011 | First experimental demonstration of ITP-aided enhancement of nucleic acid hybridization |
| Bercovici et al.[63] | 2011 | First quantitative demonstration of reaction acceleration using ITP |
| Bercovici et al.[52] | 2012 | First analytical model of ITP reaction assays, framework for several future models |
| Bahga et al.[64] | 2013 | Coupling ITP-based reaction with capillary electrophoresis for clean-up of excess reactants |
| Eid et al.[65] | 2013 | Inline reaction-separation assay using ionic spacer with sub-picomolar detection limit |
| Rubin et al.[58] | 2014 | Analytical model focusing on sample distribution within ITP zones, and its effect on reaction rates |
| Shintaku et al.[59] | 2014 | First demonstration of ITP hybridization in bead-based assays |
| Schwartz and Bercovici[69] | 2014 | First work to use ITP to speed up reactions with whole cell reactants |
| Eid et al.[57] | 2015 | Recruiting non-focusing protein reactant into ITP using modified aptamer probes |
| Eid et al.[53] | 2016 | Analytical modeling of influx and reaction rates in ITP as a function of initial sample placement |
| Phung et al.[72] | 2017 | Coupling ITP with FISH for detection of whole bacterial cells |
| **ITP to preconcentrate, mix, and accelerate heterogeneous reactions** | | |
| Garcia-Schwarz and Santiago[77,78] | 2012 2013 | ITP reaction coupled with gel-based excess reactant removal |
| Karsenty et al.[73] | 2014 | First experimentally-validated model for ITP-aided surface hybridization |
| Han et al.[74] | 2014 | Demonstration of ITP-aided DNA microarray hybridization with sub-picomolar limit of detection<br>First demonstration of truly multiplexed ITP detection |
| Shkolnikov and Santiago[76,79] | 2014 | Coupling ITP preconcentration to affinity chromatography purification<br>Comprehensive and highly predictive analytical modeling of ITP-aided capture kinetics |
| Khnouf et al.[83] | 2014 | Demonstration of ITP enhancement for surface-based immunoassays |
| Moghadam et al.[85] | 2015 | First demonstration of ITP accelerating lateral flow assays |
| Paratore et al.[75] | 2017 | Lowest limit of detection achieved by an ITP immunoassay to date |

**Table 2.** Summary of reacting species, reaction characteristics, and assay performance for the experimental studies discussed in this review. L = labeled (fluorescent, colorimetric, etc.) and i = immobilized (surface, gel, etc.)

| Authors | Species | Reaction type ($k_{on}$, $k_{off}$, $K_D$ if known) | Species controlling completion time | Reaction time | Sensitivity |
|---|---|---|---|---|---|
| Kawabata et al.[46] Park et al.[47] Kagebayashi et al.[48] | Probe 1: DNA anti-AFP WA1 antibody Probe 2: anti-AFP WA2 antibody (L) Target: Alpha feto protein | Homogeneous | Probes 1 and 2 | 2 min | 1 pM |
| Persat and Santiago[61] | Probe: Molecular Beacon (L) Target: miRNA | Homogeneous | Probe | 2 min | 10 pM |
| Bercovici et al.[63] | Probe: Molecular Beacon (L) Target: 16S rRNA | Homogeneous ($k_{on}$ = 4.75 x $10^3$ $M^{-1}s^{-1}$) | Probe | | 30 pM |
| Bahga et al.[64] | Probe: Molecular Beacon (L) Target: ssDNA oligo | Homogeneous | Probe | 3 min | 3 pM |
| Eid et al.[65] | Probe: ssDNA oligo (L) Target: ssDNA oligo | Homogeneous | Probe | 5 min | 230 fM |
| Shintaku et al.[59] | Probe: Beads conjugated with ssDNA oligo Target: ssDNA oligo | Homogeneous ($k_{on}$ = 4.4 x $10^{-6}$ $M^{-1}s^{-1}$ $K_D$ = 7.3 x $10^{-12}$ M) | Probe | 20 min | 100 fM |
| Schwartz and Bercovici[69] | Probe: antimicrobial peptides (L) Target: *E. coli* cells | Homogeneous | Probe | Up to 1 hr | 2 x $10^4$ cfu/mL |
| Eid et al.[57] | Probe: SOMAmer (L) Target: C-reactive Protein | Homogeneous ($K_D$ = $10^{-9}$ M) | Probe | 5 min | 2 nM |
| Phung et al. | Probe: ssDNA (L) Target: *E. coli* and *P. aeruginosa* cells | Homogeneous | Probe | 30 min | 6 x $10^4$ cfu/mL |
| Garcia-Schwarz and Santiago[77,78] | Reporter: Hairpin oligo Probe: ssDNA oligo (i) Target: let-7a miRNA | Gel | Reporter and probe | 15 min | 2 pM 10 pM |
| Karsenty et al.[73] | Probe: Molecular beacon (i) | Surface | Target | 3 min | 1 nM |

| | Target: ssDNA oligo | | | | |
|---|---|---|---|---|---|
| Han et al.[74] | Probe: ssDNA oligo (i)<br>Target: ssDNA oligo | Microarray (Surface)<br>($k_{on}$ = 76 mol$^{-1}$s$^{-1}$<br>$k_{off}$ = 4.4 x 10$^{-5}$ s$^{-1}$<br>$K_D$ = 5.7 x 10$^{-10}$ M) | Target | 30 min | 100 fM |
| Shkolnikov and Santiago[79] | Probe: ssDNA oligo (i)<br>Target: ssDNA oligo | Affinity column<br>($k_{on}$ = 1.5 x 10$^3$ M$^{-1}$s$^{-1}$<br>$K_D$ = 10$^{-11}$ M) | Probe | 10 min | -- |
| Khnouf et al.[83] | Probe: Antibody-coated magnetic beads (i) or antibodies (i)<br>Target: Bovine serum albumin | Surface | Probe | 2 min | 18 pM |
| Moghadam et al.[85] | Probe: rabbit anti goat IgG<br>Target: goat anti-rabbit IgG (L) | Surface<br>($k_{off}$ = 1.75 x 10$^{-3}$ s$^{-1}$<br>$K_D$ = 1.42 x 10$^{-4}$ M) | Probe | 7 min | 0.7 nM |
| Paratore et al.[75] | Probe: anti-GFP antibody<br>Target: EGFP antigen | Surface<br>($k_{on}$ = 2.1 x 10$^5$ M$^{-1}$s$^{-1}$) | Target | 6 min | 220 fM |

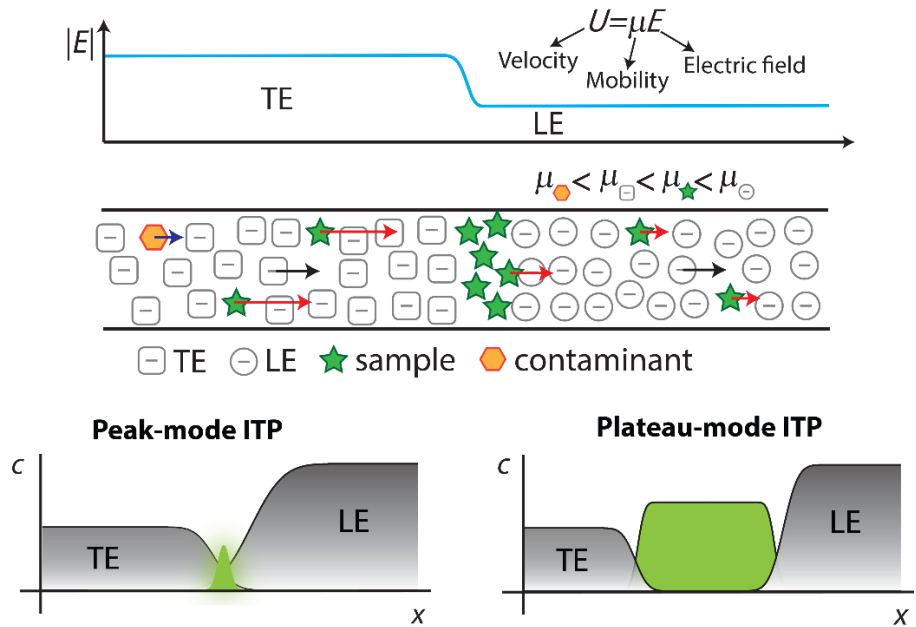

**Figure 1.** *Top:* Schematic representation of selective focusing in ITP. Sample species with intermediate mobilities migrating in the TE zone overspeed neighboring TE ions, while those migrating in the LE zone are overtaken by neighboring LE ions. Sample species therefore focus at the TE-LE interface, where their velocity will match that of the LE and TE zones (the so-called ITP velocity). Species that have a mobility lower than that of the TE electromigrate but fall behind the ITP interface, whereas species with higher mobility than the LE overtake the LE. *Bottom:* Schematic of peak and plateau ITP modes. In the former, dilute sample ions focus in a Gaussian-like peak. Multiple sample ions can co-focus in partially or entirely overlapping peaks, depending on their relative mobilities. The typical ITP peak magnitude is exaggerated considerably for visualization purposes. In plateau mode, sample ions at sufficiently high concentration to form plateaus of constant (in time) and locally uniform concentration, and contribute significantly to local conductivity.

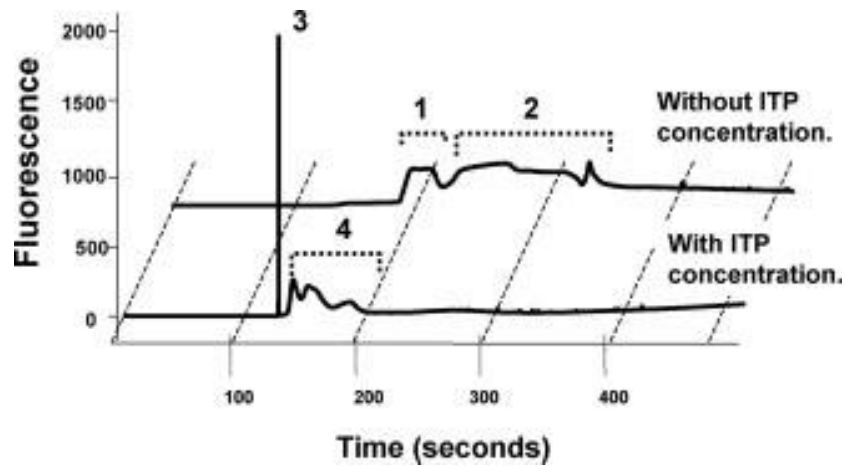

**Figure 2.** Electropherogram from Kawabata et al.[46] using of ITP preconcentration and DNA-labeled antibodies to control an ITP assay for AFP, a protein target. Peaks 1 and 3 correspond to the DNA-antibody-AFP complex, whereas peaks 2 and 4 correspond to unreacted antibodies, trailing the immune complex. ITP resulted in a 140-fold increase in fluorescent signal, enabling lower detection limits.

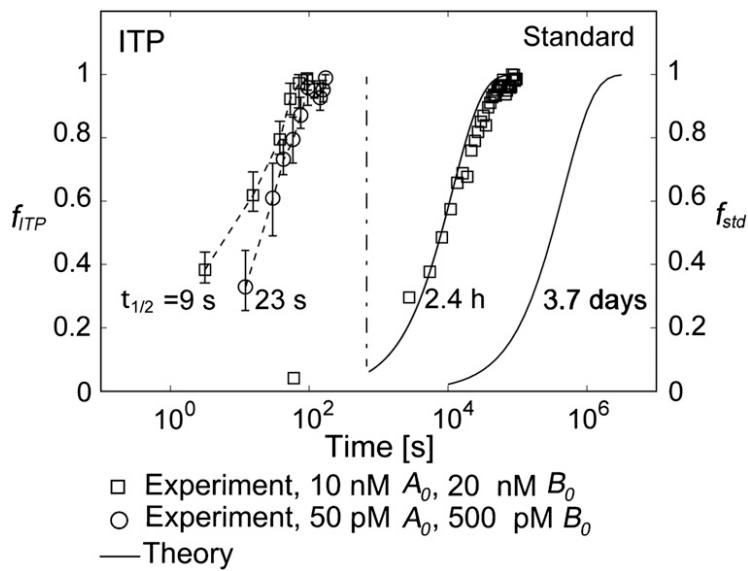

**Figure 3.** Experimental demonstration of ITP-aided hybridization acceleration from Bercovici et al.[52] Fraction of reactants hybridized is shown for both standard and ITP-aided hybridizations at two different reactant concentrations. 960- and 14,000-fold hybridization acceleration is demonstrated for limiting species concentrations of 10 nM and 100 pM, respectively. The theoretical model (discussed in Section III.a) agreed with experimental data, and captured the significant trends resulting from ITP-aided hybridization. This analytical model of ITP was the first of its kind and served as the foundation for several subsequent analytical studies.

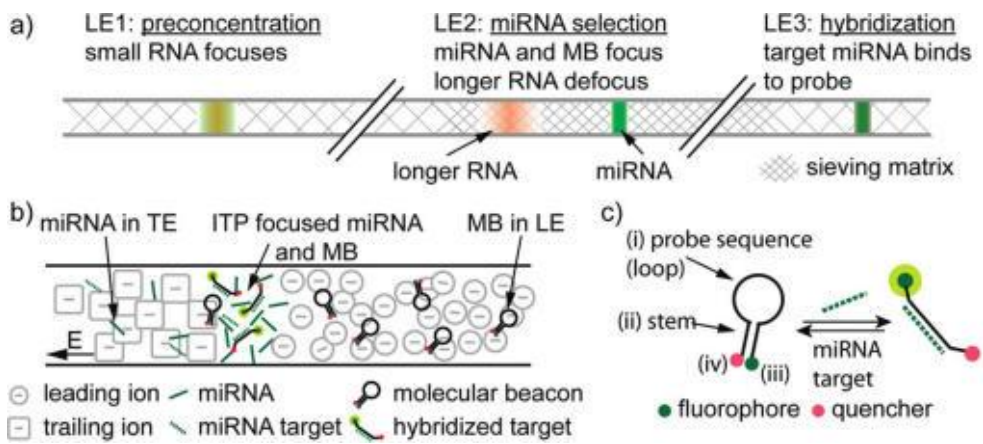

**Figure 4.** Schematic representation of the first experimental ITP hybridization assay from Persat and Santiago.[61] The assay used molecular beacons, which are fluorescently (imperfectly) quenched in their native state but unravel and emit higher fluorescence upon binding to their target. Sieving matrix was used to divide the channel into three regions to respectively promote ITP preconcentration, size selectivity, and highly stringent hybridization. This assay was able to selectively detect mature miRNA target in tissue total RNA samples, and is the first of many ITP hybridization studies to use molecular beacons as probes.

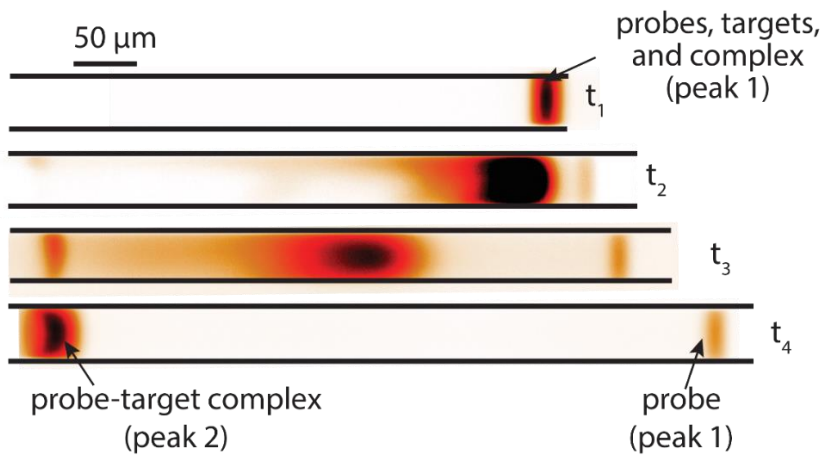

**Figure 5.** Experimental visualization from DNA reaction and separation assay using ITP and ionic spacers.[65] The probe is a 26 nt fluorescently-labeled ssDNA oligo and target a 149 nt ssDNA oligo. In the first stage ($t_1$), all reactants and products were co-focused in an ITP zone, which promoted rapid hybridization. In the second stage ($t_2$ and $t_3$), the ITP zone entered a region with sieving matrix, causing the ionic spacer to overspeed the larger probe-target complex but not the unreacted probe molecules. By $t_4$, all species had reached equilibrium, and the two ITP peaks migrated in parallel.

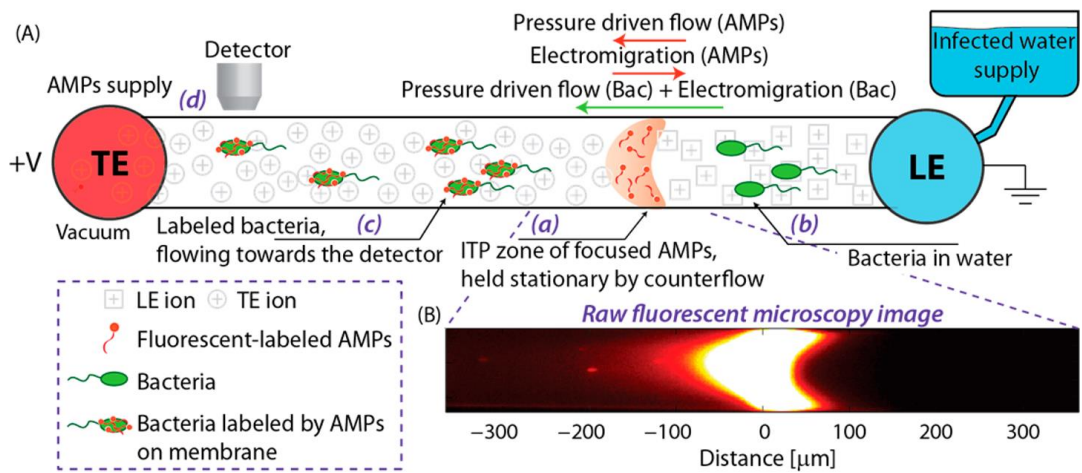

**Figure 6.** Schematic and experimental visualization of the first assay to use ITP to accelerate reactions with whole dispersed cells. Positively-charged antimicrobial peptides are focused in ITP then held stationary using counterflow, while whole bacterial cells migrate through the channel by pressure-driven flow. Peptides label the flowing cells, which are then detected downstream using a fluorescent detector. The experimental set-up was stable over the 1 h detection window.

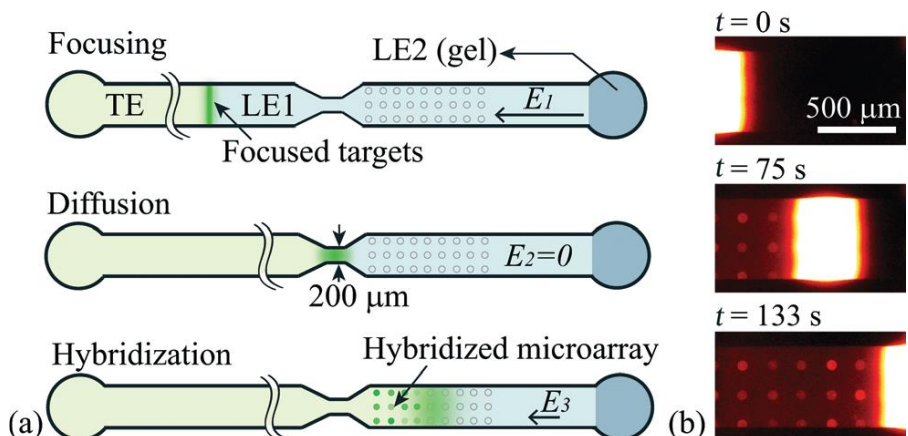

**Figure 7.** Schematic of the ITP-based multiplexed microarray assay designed by Han et al.[74] First, labeled ssDNA targets were focused in ITP. Upon reaching the constriction, the electric field was turned off to minimize Joule heating and allow redistribution of target DNA through diffusion. The ssDNA targets then migrated over 60 immobilized probe sites representing 20 unique sequences, and fluorescence was measured after the ITP zone had swept by. The 30 min assay achieved a 100 fM limit of detection, the lowest for any heterogeneous ITP reaction assay. The array signal was quantified using a standard DNA array scanner.

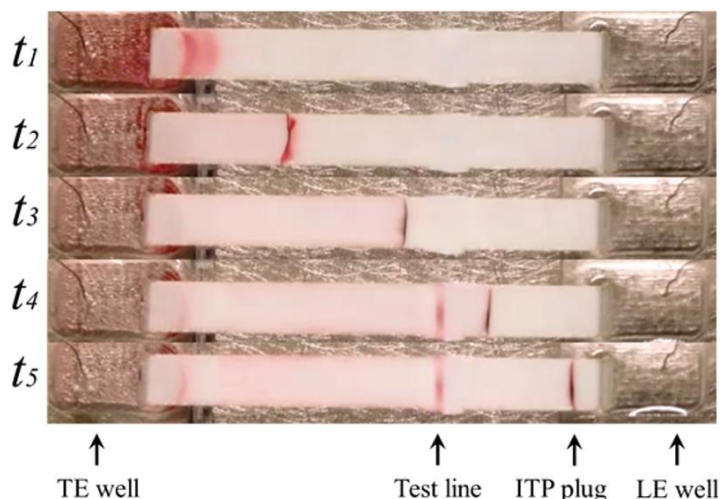

**Figure 8.** Illustration of the various stages of the ITP-LFA designed by Moghadam et al.[85] By using ITP to preconcentrate target molecules, ITP-LFA addresses the poor detection limits typical of conventional LFA. The assay demonstrated a two order of magnitude improvement in detecting IgG, and maintained the short assay times (5 min) typical of LFA.